\documentclass[12pt,english]{article}
\usepackage[T1]{fontenc}
\usepackage[latin9]{inputenc}
\usepackage{geometry}
\usepackage{amsmath}
\usepackage{amssymb}
\usepackage{babel}
\usepackage{graphicx}
\begin{document}

\title{Ideal quantum glass transitions: many-body localization without quenched disorder}

\author{M. Schiulaz\\
\textit{\small{International School for Advanced Studies (SISSA), via Bonomea 265, 34136 Trieste, Italy}}\\
M. M\"uller\\
\textit{\small{The Abdus Salam International Center for Theoretical Physics, Strada Costiera 11, 34151 Trieste, Italy}}
}

\date{}

\maketitle

\begin{abstract}
We explore the possibility for translationally
invariant quantum many-body systems to undergo a dynamical glass transition,
at which ergodicity and translational invariance break down spontaneously, driven entirely by quantum effects.
In contrast to analogous classical systems, where the existence of
such an ideal glass transition remains a controversial issue, a
genuine phase transition is predicted in the quantum regime. This
ideal quantum glass transition can be regarded as a many-body localization
transition due to self-generated disorder. 
Despite their lack of thermalization, these disorder-free quantum glasses do not possess an extensive set of local conserved operators, unlike what is conjectured for many-body localized systems with strong quenched disorder. 
\end{abstract}

\section{Introduction}

A single quantum particle placed in a sufficiently strong disorder potential does not explore the full phase space at a given energy, but remains confined in a finite spatial region, a phenomenon well-known as Anderson localization\cite{Anderson}. The absence of diffusion, and thus non-ergodicity in phase space, is particularly striking in low dimensions $d\leq 2$ where it occurs even in arbitrarily weak disorder potentials. 
A similar absence of transport and broken ergodicity due to quantum interference was predicted in many-body systems of finite density, provided that there is sufficiently strong disorder and that the interactions are sufficiently weak and short range\cite{AndersonFleishman, BAA,Gornyi,BerkovitsShklovskii, HuseOganesyan, PalHuse}. The complete absence of transport and diffusion at non-zero temperature has been termed "many-body localization".  

Both in the single particle case, as well as in the many-body systems studied so far (weakly interacting, disordered fermions, disordered bosons, random quantum magnets), quenched disorder is of paramount importance, as it stabilizes the localized phase. The disorder ensures that local rearrangements are typically associated with significant energy mismatches, which appear as large denominators in perturbation theory. Those suppress the higher order decay processes that would be necessary to establish transport and delocalization in an isolated  system that is not in contact with a thermal bath. 
 
In this article, we argue however that quenched disorder is not a necessary prerequisite to localize a many-body system. Instead we show that {\em self-generated} disorder (as a consequence of random, spatially heterogeneous initial conditions) can be sufficient to induce broken ergodicity and absence of diffusion in a closed quantum many-body system, even if its Hamiltonian is perfectly disorder-free and translationally invariant. In other words, we  predict the possibility of an {\em ideal quantum glass transition}, e.g., upon tuning the kinetic energy of some of the systems constituents. This should be compared with classical glass forming systems at finite temperature (such as, e.g., polydisperse Lennard-Jones mixtures, or hard sphere liquids), where it remains a longstanding and unresolved question whether the "glass transition" is merely a dynamic crossover or a genuine dynamic phase transition~\cite{Berthier}. In contrast, in the quantum systems analyzed here the extra ingredient of quantum interference allows us to argue that certain systems without quenched disorder do undergo a genuine dynamic phase transition towards a glass phase characterized by infinitely long-lived spatial inhomogeneities and vanishing diffusivity.

In the theory of structural glasses, especially in what is now known as the theory of random first order transitions, the close similarity between mode coupling equations for disorder-free structural glasses and the dynamics of certain mean field spin models {\em with} quenched disorder~\cite{WolynesThirumalai} was taken as a strong hint that there might be a genuine phase transition in structural glasses paralleling the glass transition in the $p$-spin model. This analogy also fostered the idea that frustrating disorder may be generated  dynamically in an otherwise disorder-free system.
While large scale spatial and temporal heterogeneities are undoubtedly present in such systems, see, e.g., the review~\cite{Berthier}, so far it has nevertheless remained impossible to prove the existence of diverging length and time scales as one crosses the putative glass transition in finite dimensional systems. The only, but somewhat trivial exception is the jamming transition at zero temperature, where systems with hard core potentials are compressed to such an extent that all particles become stuck and the system becomes incompressible and infinitely viscous.   
Given the still unclear status of the glass transition in classical systems, it appears very appealing that within the quantum world the additional ingredient of quantum interference provides a physical mechanism for complete localization in phase space and a concomitant divergence of transport and relaxation times at a quantum glass transition, even at finite temperature or energy densities. 
 
We should emphasize that with the term "quantum glass" we refer here to a system with non-ergodic dynamics, which is however perfectly closed, that is, absolutely isolated from external sources of noise, such as thermal baths. We thus employ the same notion of "glass" as sometimes used in the context of non-interacting Anderson insulators ("Fermi glass"), or for disordered bosons ("Bose glass"), even though the latter term is usually only meant to imply a bad conductor, but not necessarily the complete absence of transport. We emphasize, that these "quantum glasses", like Anderson localization in general, are not robust to thermal noise. Such external perturbations introduce dephasing and usually restore ergodicity and transport, albeit with high resistance at low temperatures. Robustness of glassiness and non-ergodicity against thermal noise requires the existence of barriers, which become arbitrarily large as the distance between the considered configurations increases, or very complicated pathways between different parts of configuration space (such as those required for the relaxation of defects in certain topological quantum systems at $T\to 0$~\cite{topglass}). The former happens in systems which undergo spontaneous symmetry breaking, as well as in many disordered systems with frustrated interactions, such as spin glasses. 
It also happens in mean-field like lattice models without disorder~\cite{BiroliMezard}. Many such models can then be endowed with additional quantum fluctuations~\cite{Chamon, Zamponi,Carleo,Nussinov}, which then naturally inherit the glassiness present due to the classical frustration, as long as quantum and thermal fluctuations do not destroy the glassy order. While these models exhibit interesting effects arising due to quantum fluctuations (reentrant transition lines, discontinuous glass transitions at low $T$, etc), the role of quantum effects is secondary to the extent that the glassiness of these systems relies fundamentally on the built-in frustration in their classical configuration space.   
In this paper we study a very different type of quantum glass, in which classical frustration is absent, whereas quantum interference and the Anderson localization mechanism are the {\em only} driving forces that entail ergodicity breaking. 
 
\section{Ingredients of ideal quantum glasses}

Let us first discuss the elementary ingredients necessary to obtain an ideal quantum glass as envisioned above. In the subsequent section, we will make these concepts explicit by discussing in detail a simple one-dimensional model. 
We consider Hamiltonians which are the sum of two parts,
\begin{equation}
H=H_{0}+\lambda T.\label{H}
\end{equation}
$H_0$ describes the "non-hopping part" of the many-body system, of which require that dynamics under $H_0$ is trivially non-ergodic and localized, and by itself would not allow for any d.c. transport. 
In this sense, $H_0$ is the analogue of the disorder potential in a standard single particle Anderson problem.
In the context of non-disordered, translationally invariant Hamiltonians, it is important to note that the absence of hopping leads to an extensive degeneracy of the spectrum, which allows one to choose the highly degenerate eigenbasis of $H_{0}$ in the form of localized many-body wavefunctions, which break translational invariance. For our purpose this will be the natural basis from which we will construct eigenstates of the full Hamiltonian (\ref{H}), rather than to use a basis which respects translational invariance.

Consider now adding a perturbative "hopping part"  $\lambda T$ to the Hamiltonian. We choose it such that it formally could restore ergodicity, in the sense that any state in Hilbert state can be reached from any eigenstate of $H_0$ by the successive action of appropriate terms appearing in $T$ (in analogy to the intersite hopping in the Anderson problem, which in principle could bring a particle anywhere in the lattice). Our aim is to argue that, for $\lambda$ sufficiently small, the system nevertheless remains non-ergodic and localized, while for large enough $\lambda$ a dynamical quantum glass  transition into an ergodic quantum liquid is expected. 

By nonergodic we mean that generic, microscopically inhomogeneous initial conditions will not become more and more homogeneous with increasing time; but rather that equilibration is avoided, even locally, and inhomogeneity persists forever. More precisely, for typical local observables $O(r)$, such as energy density or particle density, the long time limit 
\begin{equation}
\lim_{T\to \infty}\frac{1}{T} \int_0^T dt\, \left[\lim_{L\to \infty} \langle O(r,t)\rangle \right]\neq \langle\langle O\rangle\rangle_{\rm th}\,,
\label{nonergodicity}
\end{equation}
will not tend to its space-independent equilibrium value. Note that it is important to take the thermodynamic limit before the infinite time limit. In finite size samples the long time dynamics will be similar to that in the thermodynamic limit up to times that are exponentially large in the system size. However, beyond those times, very slow transport and diffusion may be observed and translational invariance will be restored upon averaging over those large time scales. Nevertheless, local observables and correlation functions probably still do not tend tend to their equilibrium values in the regime of parameters where a localized phase is expected in the thermodynamic limit.  


An important element in establishing the non-ergodicity and persistence of inhomogeneity anticipated in Eq.~(\ref{nonergodicity}) consists in showing that eigenstates remain close to the spatially inhomogeneous eigenstates of the system at $\lambda=0$. In particular, we will argue that the full eigenstates spontaneously break translational invariance in the thermodynamic limit. Therefore  they obviously  violate the eigenstate thermalization hypothesis~\cite{ETH}, which is a conjecture expected to hold for ergodic quantum systems. It considers pure states with density matrix $|\Psi\rangle\langle \Psi|$, where $\Psi$ is a typical eigenstate at given energy, momentum and particle numbers. The conjecture is that, in the thermodynamic limit, the reduced density matrix of a subsystem much smaller than the system size tends to the reduced density matrix of the Gibbs ensemble, and thus the pure state realizes a local equilibrium in any small subpart of the system. This conjecture has been pointed out to break down in many-body localized quantum systems.  

\section{An ideal quantum glass in 1d}
\subsection{Inhibited hopping model}
\begin{figure}
  \includegraphics[width=.8\textwidth]{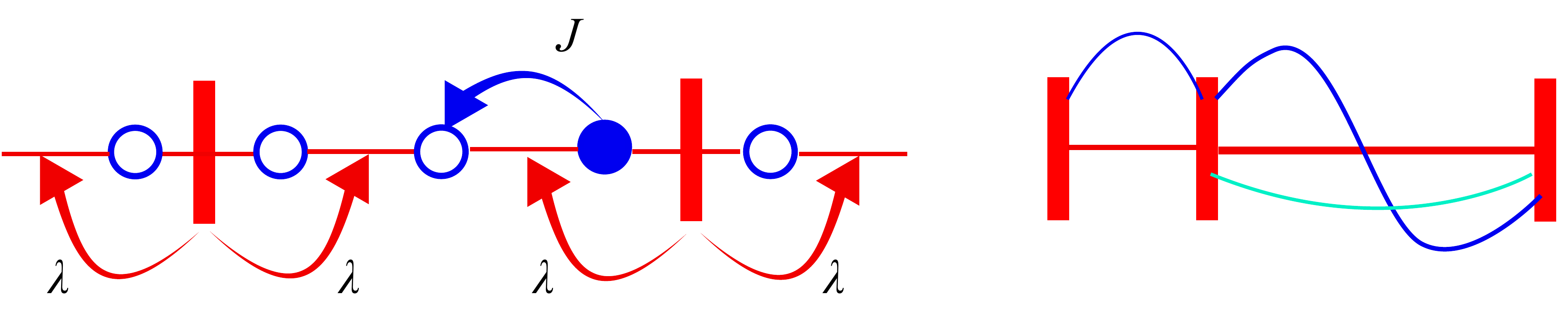}
  \caption{(Color online) Left: A simple 1d model exhibiting self-induced many-body localization: Barriers (red) that live on the links of the chain block the propagation of fast particles (blue). Since $J\gg \lambda$, the fast particles produce a random energy landscape for the barriers, which  Anderson localize as a consequence. - Right: The non ergodic Hamiltonian $H_{0}:$ the barrier divide the chain into independent intervals, where the fast particles occupy free particle states.}
  \label{fig:model}
\end{figure}

To exemplify and analyze the general phenomenon of non-disordered, self-localizing quantum glasses, we consider a simple 1d model containing two kinds of fermions: a "fast" species $a$ and a "slow" species $b$. The presence of slow particles is assumed to hinder the propagation of the fast particles, cf. Fig.~\ref{fig:model}. Physically, this may arise due to a strongly enhanced tunneling barrier for a fast particle in the presence of a $b$ particle, which we therefore call a "barrier" henceforth. The Hamiltonian we consider takes the form of \eqref{H}, with  
\begin{eqnarray}
\nonumber
H_{0}&=&-J\sum_{i}\left(a_{i+1}^{\dagger}a_{i}+a_{i}^{\dagger}a_{i+1}\right)\left(1-b_{i}^{\dagger}b_{i}\right),\\
\lambda T&=&-\lambda\sum_{i}\left(b_{i+1}^{\dagger}b_{i}+b_{i}^{\dagger}b_{i+1}\right).
\label{Model}
\end{eqnarray}

The barriers move with very small kinetic energy $\lambda \ll J$ as compared to the hopping strength $J$ of the fast particles. Note that in the hopping part $T$, we could also add a hopping of $a$ in the presence of a barrier, without altering the qualitative conclusions of the analysis below.

The index $i=1,...,L$ labels both the lattice sites that host the fast particles, and the links between them where the barriers reside, cf. Fig~\ref{fig:model}. Further, we impose periodic boundary conditions, identifying sites $1$ and $L+1$, so as to ensure the full translational invariance of the Hamiltonian.
We fix the number of particles $N_{a,b}$, and their respective densities $\rho_{a,b}=N_{a,b}/L$.
This model is somewhat reminiscent of Falicov-Kimball models, where localized particles create a potential for inert fast particles. However, here we are not interested in the thermodynamic properties, such as the ground state, but rather in dynamic questions: giving the  heavy particles a finite mass, we ask about the coherent dynamics of the whole system and its ergodicity and transport properties.

\subsection{Properties of the non-hopping Hamiltonian $H_0$} 
For $\lambda=0$, the barriers cannot move. Hence, the fast particles remain confined between them and do not interact with other particles outside their own interval. Thus, there is trivially no long range transport of density or energy, and the system is strongly non-ergodic.  Each interval between two consecutive barriers hosts a {\em discrete} set of levels for the fast particles. In the simple model above, for intervals of length $\ell$, these are just standing waves with wavevectors $k_m =  m\pi/(\ell+1)$, for $m=1,...,\ell$, wavefunctions $\psi^{(\ell; m)}_j = \sqrt{\frac{2}{\ell+1}} \sin\left(k_{m}j \right)$ and energy $E_{\ell;m} = -2J\cos(k_{m})$, but these specific forms are inessential for our subsequent considerations, except for the fact that the level spacings within a given interval are of order $O(J)$ and depend on $\ell$. In fact we explicitly neglect accidental degeneracies of different level spacings, as they are non-generic and can be removed by simple modifications of the Hamiltonian for the fast particles. 
Under these circumstances, a barrier can only move with a concomitant energy change in the spectrum of the fast particles, unless the barrier motion does not alter the distribution of interval lengths. 
If one neglects that latter exception, one realizes that fast particles placed in a random arrangement of barriers create an inhomogeneous energy landscape for the barriers, which plays the role of disorder in standard non-interacting and interacting localization problems. At first sight it thus appears almost obvious that when barriers acquire a small but finite hopping $\lambda$, the latter cannot compete with the much larger roughness in the energy landscape, and thus the motion of the barriers becomes fully localized. However, this view, even though basically correct, is oversimplified and misses several potential caveats. First, the non-hopping Hamiltonian $H_0$ has an extensively degenerate spectrum. This is so because any permutation of the intervals (as defined by consecutive barriers) does not cost any energy at $\lambda=0$ if the internal state of fast particles is permuted together with the intervals. Thus, one has to be careful in dealing with these resonances before one can assert self-induced localization.
Secondly, it should be clear that breaking of translational invariance can only occur in the thermodynamic limit, since it is easy to show that, in any finite system with periodic boundary conditions, the long time average must necessarily be translationally invariant. Hence, the limit of infinite volume should be carefully discussed. Finally, we will only be able to argue for localization of the most abundant, typical initial conditions, while we will see that initial conditions with extensive correlations are not localized.  This implies that in our model localized states coexist with very rare delocalized states in the same energy range. Numerical results suggest that this might possibly also be the case in many body localized systems with quenched disorder~\cite{Bauer}. 
The non-trivial aspect of our analysis consists thus in showing that the extensive degeneracy of $H_0$ is lifted by introduction of the hopping in such a way that strong resonances and system-spanning hybridization are avoided for {\em typical}, random initial conditions. 
  
From the above considerations we thus expect a dynamical transition between two phases: at small $\lambda$, diffusion is prevented, the dynamics of the system are non-ergodic and the eigenstates of the Hamiltonian remain localized in the many-body Hilbert space (to be defined more precisely below). As we will show, in this regime, translational invariance of the eigenstates is spontaneously broken.
In contrast, when $\lambda$ becomes large enough, many configurations of the non-hopping Hamiltonian start hybridizing and cause the many-body wavefunctions to  
delocalize throughout Hilbert space. In the time evolution from an arbitrary initial condition, one thus expects the system to relax to a state of local equilibrium with spatial homogeneity restored (in the long time average).

An essential property of $H_0$ is its local "integrability", i.e., the existence of an extensive set of mutually commuting, local conserved quantities. In the above model, apart from the trivially conserved barrier positions, $b_i^\dagger b_i$, the conserved quantities associated with the levels of fast particles take the form 
\begin{equation}
\tau_{i,\ell;m}=b_{i}^{\dagger}b_{i}\, b_{i+\ell}^{\dagger}b_{i+\ell}\,\prod_{j=1}^{\ell-1}\left(1-b_{i+j}^{\dagger}b_{i+j}\right)\, b_{i+\ell}^{\dagger}b_{i+\ell}\,\gamma_{i,l;m}^{\dagger}\gamma_{i,l;m},\quad1\leq m\leq\ell,\quad1\leq i,\ell\leq L,
\end{equation}
where $\gamma^\dagger$ are the creation operators for fast particle states confined by two barriers located at the links $i$ and $i+\ell$,
\begin{equation}
\gamma_{i,l;m}^{\dagger}= \sum_{j=1}^{\ell}\psi^{(\ell;m)}_j\,a_{i+j}^{\dagger}.
\end{equation}
With these integrals of motion, $H_0$ can be compactly rewritten as 
\begin{equation}
H_{0}=\sum_{i=1}^{L}\sum_{\ell=1}^{L}\sum_{m=1}^{\ell}E_{\ell;m} \tau_{i,\ell;m}.
\end{equation}

\subsection{Eigenstates of $H_0$ and $H$}
In a generic high energy eigenstate of $H_0$ (with energy density of order $O(J)$ above the ground state), the intervals between nearest neighbor barriers have lengths $\ell$, which are exponentially distributed according to the probability distribution
\begin{equation}
p(\ell) = (1-\rho_b)^{\ell-1}\rho_b,
\label{expdist}
\end{equation} 
with mean length $\overline{\ell}=1/\rho_b$. The fast particles trapped between the barriers occupy any of the discrete energy levels $E_{\ell;m}$ discussed above. Let us now discuss how such eigenstates deform when a finite but small hopping of barriers is turned on.

\subsubsection{Broken translational invariance} Our aim is to show that the eigenstates remain "localized" close to the eigenfunctions of $H_0$. In spirit, we follow a similar route as taken in the   approach by Imbrie and Spencer~\cite{Imbrie}, attempting to rigorously prove many-body localization in disordered quantum  Ising chains. Here, we do not aim at controlling localization with mathematical rigor, but rather focus on highlighting the additional aspects and subtleties to be considered when no quenched disorder is present.
 
 We start from the eigenbasis of $H_0$ discussed above, which breaks translational invariance explicitly. Our strategy is to show that the higher we push the perturbation theory in $\lambda$, the smaller the fraction of states which remain degenerate at the given order of perturbation theory. Nevertheless, the degeneracy associated with translations always survives. For typical, that is, random eigenstates, that degeneracy will only be lifted at a perturbative order proportional to the system size $L$.  All other degeneracies will be lifted fairly rapidly, and persist typically at most to an order that grows logarithmically with $L$, due to rare subsequences of barrier intervals. 
 
The extensive degeneracy between eigenenergies is lifted at various levels. In general, we need to use degenerate perturbation theory. Let us consider the eigenstate $\left|\chi_{m}^{\left(0 \right)}\right\rangle$ of $H_0$. At the $k$-th order of perturbation theory, we consider all the  states $\left|\chi_{n}^{\left(k-1 \right)}\right\rangle$, constructed at the previous stage, which have matrix elements with $\left|\chi_{m}^{\left(k-1 \right)}\right\rangle$ at this order.
If the relevant matrix element is much smaller than the energy
difference between the two states (as obtained at $(k-1)$-th order), we consider the state \textit{off-resonance}
with $\left|\chi_{m}^{\left(k-1\right)}\right\rangle $ and treat it
using non-degenerate perturbation theory. If, in contrast, the matrix
element is dominant, it is necessary to write an effective Hamiltonian
in the resonant subspace, with matrix elements of order $O(\lambda^{k})$
as non-diagonal elements,
and the energies evaluated up to order $O(\lambda^{k-1})$
as diagonal entries. 

The core of our argument consists in showing that by applying perturbation
theory to a random eigenstate of $H_{0},$ its random barrier configuration ensures the lifting of degeneracy at low perturbative orders. The degeneracy, which is present in $H_0$ due to invariance of the  energy under permutation
of interval lengths, is lifted due to perturbative shifts in the eigenenergies, which are sensitive to the actual sequence of intervals. The goal is to show that, 
when at higher order the initially degenerate configurations are connected by a non-zero matrix
element between each other, they are no longer in resonance, since
their degeneracy has already been lifted by an amount which parametrically exceeds the typical matrix element. As we will show, for small $\lambda$ resonances become rapidly rarer
as the perturbative order is increased. Thus, perturbation theory will converge, and the full eigenstate $\left|\chi_{m}\right\rangle$
remains close to $\left|\chi_{m}^{\left(0\right)}\right\rangle$. 
By the latter, we mean that expectation values of local observables (such as the position of barriers) remain close to their values on the unperturbed states - except in rare spatial regions where significant hybridizations take place. Note that this notion of closeness does not require that $\left\langle \chi_{m} |\chi_{m}^{\left(0\right)}\right\rangle$ is finite; indeed such an overlap trivially decays exponentially with system size, for any finite $\lambda$. The important point is rather that $\left|\chi_{m}^{(0)}\right\rangle$ hybridizes with far less states in Hilbert space than a wavefunction would do whose pure state would realize a local Gibbs ensemble.
 
\subsubsection{Dynamical consequences of the localization of many-body eigenstates}
The convergence of the perturbation theory for typical eigenstates implies that they can still  be characterized essentially by the position of barriers, since their expectation values will still remain strongly peaked at the sites where they were in the state  $\left|\chi_{m}^{\left(0\right)}\right\rangle$. This has important consequences on the dynamics. Namely, consider the initial value problem where the system is prepared in a certain random configuration of barrier positions and arbitrary states of fast particles in between. Formally, the dynamical problem can be solved by an expansion into eigenstates, whose time evolution is simple. The important point to observe is that the weight of eigenstates  appearing in this decomposition and having average barrier positions that significantly differ from that of the initial condition in $N$ sites, decreases exponentially with $N$. From this we conclude that the probability for $N$ barriers to move substantially away from their initial position simultaneously is exponentially small. Thus the system is definitely non-ergodic since spatial inhomogeneities persist forever. Similarly, one can argue that transport of energy and particles is absent in almost all initial configurations. This is because an initial inhomogeneity, e.g., in the of density of barriers, cannot relax, and thus diffusion must be absent.    
 
\subsubsection{Lifting of degeneracies}
\begin{figure}[h]
\includegraphics[height=.2\textheight]{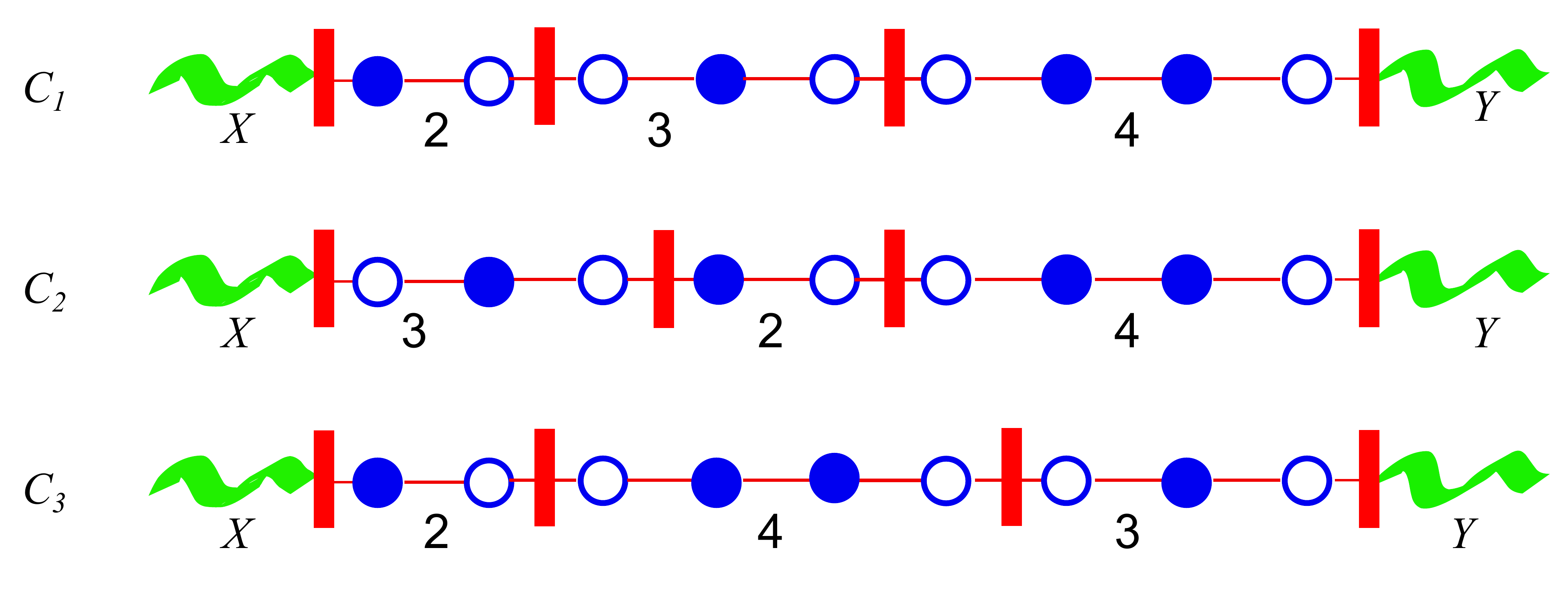}
\caption{Barrier configurations (eigenstates of $H_0$) that hybridize at first order of perturbation theory in $\lambda$. Every configuration is related to at least one other configuration by moving one barrier by one lattice spacing, exchanging the adjacent lengths of intervals between $(\ell, \ell+1)$ and $(\ell+1, \ell)$.  
The occupation of levels of fast particles in the intervals with equal length must be the same for the states to be degenerate.}
\label{fig:1st_order_deg}
\end{figure}

The extensive degeneracy between eigenenergies is lifted at various levels. At first order in perturbation theory the only degenerate eigenstates of $H_0$ that can be connected by $\lambda T$ are configurations where one barrier moves such that the two adjacent intervals exchange lengths from $(\ell, \ell+1)$ to $(\ell+1, \ell)$. This is illustrated in  Fig.~\ref{fig:1st_order_deg}, which shows a small part of an extended system, with barriers arranged in the following patterns of interval lengths:
 \begin{equation}
\left|C_{1}\right\rangle =\left|X,2,3,4,Y\right\rangle,
\end{equation}
where $X$ and $Y$  denote the unspecified pattern of barriers to the left and the right of the considered subsystem.
The configuration of fast particles is not specified, but must always be the same within   intervals of equal lengths, so as to ensure degeneracy at $0^{\rm th}$ order.
At first order in $\lambda,$ this state is connected to the following degenerate configurations shown in Fig.~\ref{fig:1st_order_deg}:
\begin{eqnarray*}
\left|C_{2}\right\rangle&=&\left|X,3,2,4,Y\right\rangle \quad {\rm and }\quad
\left|C_{3}\right\rangle=\left|X,2,4,3,Y\right\rangle .
\end{eqnarray*}
The off-diagonal elements $T_{ij}\equiv\left\langle C_{i}\right|T\left|C_{j}\right\rangle$ are non-zero for $(ij)= (12)$ and $(13)$, which induces a level splitting of order $O(\lambda)$.

\begin{figure}[h]
\includegraphics[height=.15\textheight]{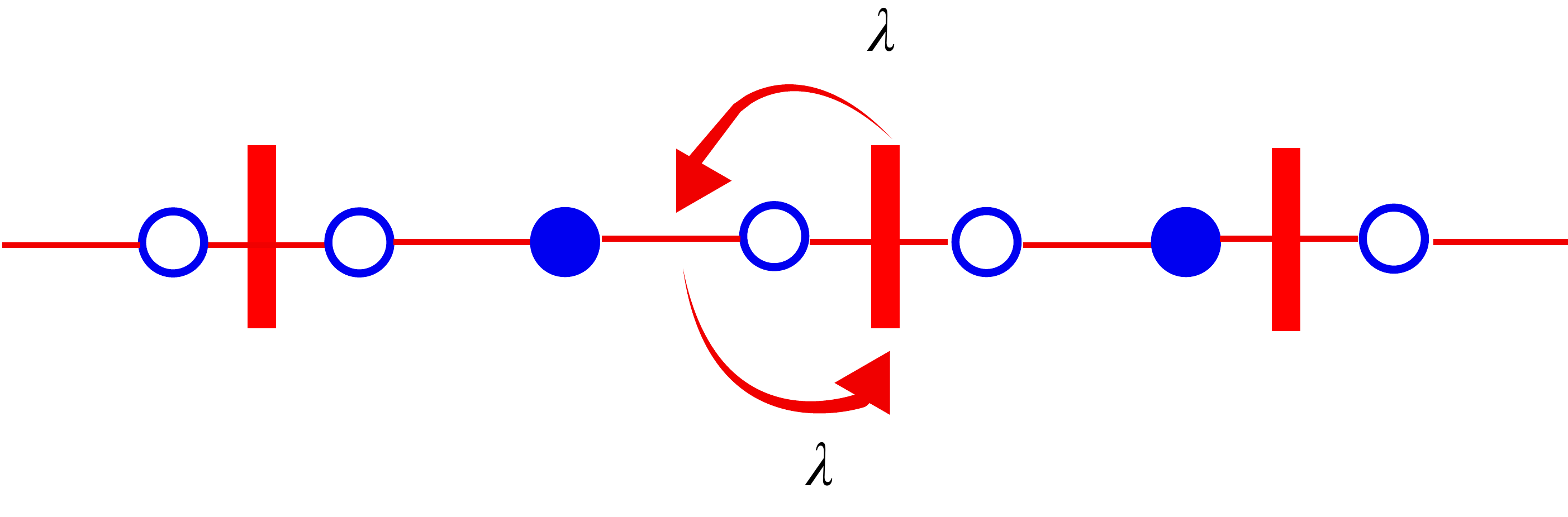}
\caption{The energies of eigenstates are shifted by contributions of order $O(\lambda^2/J)$ by the virtual hops of barriers. These virtual processes lift the largest part of the degeneracies in the many-body spectrum.}
\label{fig:2nd_order_shift}
\end{figure}

Apart from these local first order splittings, the most generic lifting of degeneracies takes place at second order in $\lambda$. It is due to barriers moving virtually back and forth, as shown in Fig.~\ref{fig:2nd_order_shift}. These processes lead to energy shifts of order $O(\lambda^2/J)$ per moving barrier and remove a large part of the exact degeneracies present at $\lambda=0$. 

\begin{figure}
\includegraphics[width=\textwidth]{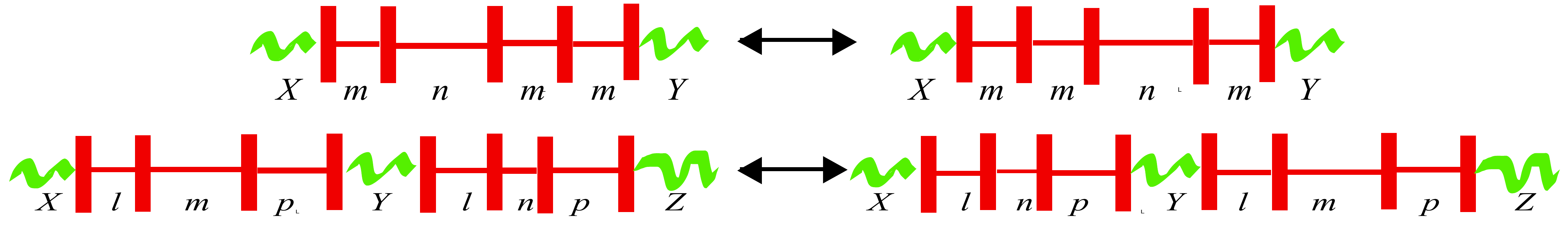}
\caption{Pairs of configurations where the exchange of an interval of length $m$ and $n$ do not change the set of pairs of adjacent intervals in the sequence. Eigenstates which differ only by such local rearrangements remain degenerate at order $O(\lambda^2)$ in perturbation theory.}
\label{fig:2nd_oder_deg}
\end{figure}

Nevertheless, certain configurations remain degenerate at higher orders of perturbation theory. These correspond to sequences of intervals, which can be permuted in such a way that each interval has neighbors of the same length as before the permutation. This ensures that the second order shifts of the energy are exactly the same in the two sequences. 
The most abundant type of configurations which are not split at second order corresponds to sequences of the form
\begin{equation}
 \left|X,m,n,m,m,Y\right\rangle ;\;\left|X,m,m,n,m,Y\right\rangle,
\label{degeneracy}
\end{equation}
as illustrated in Fig.~\ref{fig:2nd_oder_deg}. Here $m,n$ are interval lengths such that $\left|m-n\right|>2$  -- otherwise the two configurations would hybridize at second order, which would already lift their degeneracy. 
The occupation of fast particle levels within corresponding intervals of the same length is again assumed to be equal. 

In order to ensure the convergence of the perturbative expansion, the probability per unit length, that degeneracies survive, needs to be small enough and to decrease sufficiently fast with increasing order at which the degeneracy is lifted. Otherwise neighboring degenerate regions would hybridize and form a delocalized band of excitations.    
Let us therefore estimate the probability $P_{mnmm}$ of  degenerate configurations as in (\ref{degeneracy}) to appear in a given location in the sequence of intervals in a random eigenstate of $H_{0}$. Those configurations can be checked to be the most abundant type of degeneracies.The average distance between such degenerate regions will be $d_{\rm deg} \approx 1/P_{mnmm}$.
The degeneracy of these configurations is in general lifted at order $O(\lambda^{4})$. On the other hand, we expect a matrix element between two degenerate regions to appear typically at order $\lambda^{2(d_{\rm deg}-3)}$, since a matrix element is generated by the motion back and forth of the $d_{\rm deg}-3$ barriers located between the resonant regions. Therefore, if $\lambda$ is sufficiently small, such two regions will almost always be off-resonance, if $d_{\rm deg} > 5$ .  We will now show that this condition is indeed well satisfied in general. \footnote{We note that for other more complex models an analogous calculation may be more involved, and the existence of a quantum glass phase may be less evident than in the present toy model.}   

The probability $P_{mnmm}$ of either of the configurations (\ref{degeneracy}) to appear in a given location of the sequence of intervals, is easily calculated to be
\begin{eqnarray*}
P_{mnmm} & = & 
2\sum_{S=1}^{\infty}p\left(S\right)^{3}\sum_{S', |S'-S|> 2} p(S')  =\frac{2\rho_b^2}{3}+O(\rho_b^3),\end{eqnarray*}
where we explicitly use the independence of successive interval lengths.
The asymptotics $P_{mnmm} \sim \rho_b^2$ for small barrier densities $\rho_b$ arises because the first and the last interval lengths are not free, but must equal one of the two lengths in the middle. For $\rho_b=1/2$, which is nearly optimal for such degeneracies to to occur, we still find a very small probability $P_{mnmm}=0.034$, which implies a large typical separation between resonant regions by $d_{\rm deg}\approx 30\gg 5$. 

Configurations which remain degenerate at yet higher order are even rarer. The most abundant pairs whose degeneracy is not lifted at order $O(\lambda^{2n})$ correspond to sequences of intervals which can be permuted in such a way that all sets of $n+1$ consecutive intervals are present in both configurations (possibly up to spatial inversion of the sequence). As an example,  the most probable configurations that are not split at  order $\lambda^{4}$ are of the form:
\[
\left|X,l,m,n,m,m,l,Y\right\rangle ;\;\left|X,l,m,m,n,m,l,Y\right\rangle ,
\]
with the restriction $\left|m-n\right|,\left|m-l\right|>4.$ At small $\rho_b$, the density of these configurations is of order $\rho_{b}^{n+1}$. Since the inverse of this small quantity controls the power of $\lambda$ at which matrix elements couple neighboring degeneracies, hybridizations between such regions are exceedingly rare and remain strongly localized. 

However, one should be careful when reasoning about finite system sizes, where translational invariance is ultimately restored, as we mentioned above. Indeed, the degeneracy between a sequence of intervals and its rigid
translation by a certain number of lattice sites is never lifted to any order in perturbation theory.
However, when $\lambda$ is small, the matrix element connecting two such configurations is of order $O(\lambda^{\rho_{b}L})$, since all barriers need to be moved. These matrix elements give rise to an exponentially small splitting between the $L$ hybridizing, rigidly rotated configurations. Consequently, the   
characteristic time needed to observe the effect of this hybridization is inversely proportional to these matrix elements and thus diverges exponentially with the system size. Nevertheless translational invariance is restored in such finite size systems when averages are taken over times exceeding that hybridization time. These considerations are of course closely analogous to those  one makes for finite size systems undergoing the spontaneous breaking of a discrete symmetry. The relevant symmetry here is the discrete translational invariance of the lattice model. In the infinite volume limit, this remaining hybridization becomes irrelevant and spontaneous symmetry breaking occurs.

From the above arguments it becomes clear that the degeneracy of $H_0$ is lifted at higher orders in perturbation theory in the barrier hopping $\lambda$, and that quantum fluctuations induce an effective disorder, which takes the same role as quenched disorder in other many-body systems considered previously. Having reached this stage, one can repeat the same type of arguments as in those systems~\cite{BAA} to conclude that for sufficiently small $\lambda$ perturbation theory should converge and the perturbed eigenstates remain close to the inhomogeneous initial states~\cite{Imbrie}.

\subsection{Spontaneous breaking of translational symmetry of eigenstates}
In order to verify the phenomenon of self-induced many-body localization, it is useful to exploit a specific property of non-disordered systems: namely, that dynamical localization of typical quantum states, and the spontaneous breaking of translational invariance are essentially equivalent, as we argued above. From a numerical point of view this is very convenient, since the sought phenomenon can be phrased in the familiar language of spontaneous symmetry breaking. In this way we avoid the  identification of many-body localization by other observables, which are harder to analyze. Those include  the observation of freezing via Edwards-Anderson-type order parameters~\cite{PalHuse},
or the analysis of many-body level statistics~\cite{BerkovitsShklovskii,HuseOganesyan}. The latter is, however, based on the conjecture that a delocalization transition in a many-body system is concomitant with a change from Poisson to Wigner-Dyson statistics. Even if this conjecture is true and also applies to the non-disordered systems considered here,  we nevertheless would expect the corresponding  observables to suffer from stronger finite size effects than in systems with quenched disorder. Therefore, it is very useful to have an alternative route to detecting many-body localization.

In order to probe for spontaneous translational symmetry breaking, we proceed in the usual way. We introduce a small symmetry breaking term $H_{\rm SB}$ in the Hamiltonian,
\begin{equation}
H\rightarrow H+WH_{\rm SB},
\end{equation}
and ask whether the induced symmetry breaking persists as the strength of the perturbation $W$ tends to zero, after the thermodynamic limit has been taken. To break the translational symmetry externally, we apply a weak disorder potential,
\begin{equation}
H_{\rm SB}=\sum_{i}\left[\varepsilon_{i}^{a}a_{i}^{\dagger}a_{i}+\varepsilon_{i}^{b}b_{i}^{\dagger}b_{i}\right],
\end{equation}
where $\varepsilon_{i}^{a,b}$ are independent, identically distributed random variables, taken from a centered box distribution of unit width. In order to probe dynamical translational symmetry breaking, we  define for any quantum state $\Psi$ the observable
\begin{equation}
\overline{\Delta\rho}^\Psi =\frac{1}{L}\sum_{i=1}^{L}\left| \langle \Psi| b_{i+1}^{\dagger}b_{i+1}-b_{i}^{\dagger}b_{i} | \Psi\rangle \right|,
\end{equation}
which is a measure of the spatial inhomogeneity of the density of barriers. Note that for any translationally invariant state $\Psi$ (i.e., a momentum eigenstate) $\overline{\Delta\rho}^\Psi $ vanishes.
Translational invariance is present in the long time average over the dynamics, if the inhomogeneity of typical many-body {\em eigenstates} vanishes in the limit $W\to 0$. This is expected to happen if the  barriers are sufficiently mobile, i.e., for $\lambda>\lambda_c$ where $\lambda_c$ is a critical hopping strength. In contrast, translational symmetry is spontaneously broken in the dynamics starting from random (typical) initial states, if infinitesimal disorder induces a finite inhomogeneity of eigenstates in the thermodynamic limit, i.e., if
\begin{equation}
\lim_{W\to 0} \lim_{L\to \infty} \Delta\rho(\epsilon) \neq 0,
\end{equation}
where 
\begin{equation}
\Delta\rho(\epsilon) = \left\langle \overline{\Delta\rho}^{\Psi}\right\rangle_{\Psi,\epsilon}
\end{equation}
denotes an average over eigenstates with energy densities in a narrow range around $\epsilon$.
The critical value $\lambda_c$, where the quantum glass breaks down and ergodicity and transport is restored, is expected to depend on $\epsilon$, since the occupation probability of fast particle levels will affect the motion of the slow barriers. 

The effect of a weak disorder potential can be analyzed using perturbation theory in $W$. In contrast to our previous analysis, here we start with the translationally invariant eigenbasis, which simultaneously diagonalizes the momentum. 
To first order, the $n$'th eigenstate $\left|\Phi_{n}\right\rangle$ is perturbed by the following hybridizations: 
\[
\left|\Phi_{n}\right\rangle \rightarrow\left|\Phi_{n}^{W}\right\rangle =\left|\Phi_{n}\right\rangle +W\sum_{m\neq n}\frac{\left\langle \Phi_{m}\right|H_{\rm SB}\left|\Phi_{n}\right\rangle }{E_{m}-E_{n}}\left|\Phi_{m}\right\rangle+O(W^2),
\] 
where $E_{n}$ is the energy of state $\left|\Phi_{n}\right\rangle$ at zeroth order in $W$. 

Recalling that  $\overline{\Delta\rho}^{\Phi_n}=0$, we find the linear response
\begin{equation}
\overline{\Delta\rho}^{\Phi_{n}^{W}}=2W\, \textrm{Re}\left[\sum_{m\neq n}\frac{\left\langle \Phi_{m}\right|H_{\rm SB}\left|\Phi_{n}\right\rangle }{E_{m}-E_{n}}\,
\frac{1}{L}\sum_{i=1}^{L}\left|\left\langle \Phi_{n}\right|b_{i+1}^{\dagger}b_{i+1}-b_{i}^{\dagger}b_{i}\left|\Phi_{m}\right\rangle \right|\right]+O(W^{2}).
\label{linresponse}
\end{equation}

Note that here we have implicitly assumed that there are no exact degeneracies in the spectrum of the unperturbed Hamiltonian $H$, since otherwise we would have to apply degenerate perturbation theory. However, the model of Eq.~\eqref{H} is not only invariant under translations but also under spatial reflection of the lattice,
\[
\left(a_{i},b_{i}\right)\rightarrow\left(a_{L-i+1},b_{L-i+1}\right),
\]
which implies a double degeneracy of most levels in the spectrum of $H$. Such degeneracies lead to a singular response of translationally invariant eigenstates to weak disorder, which mixes the two degenerate states. This  effect makes it difficult to disentangle the spontaneous breaking of the inversion symmetry breaking from that of translational symmetry using the above probe.
To circumvent this issue, we have eliminated the inversion symmetry by applying an incommensurate magnetic flux $\phi=\pi/\sqrt{2}$ through the ring formed by the periodic chain.

To estimate the linear response we assume that it is dominated by pairs of states $\Phi_{m,n}$ which belong to the same miniband of $L$ states which consist in hybridizations of rigidly rotated, localized barrier configurations. Let us denote the $L$ localized states by $\left|j \right\rangle$, where $j=0,...,L-1$ is the shift from a reference position of the considered configuration. (We assume a typical configuration, for which all obtained configurations are different). The relevant states dominating in Eq.~(\ref{linresponse}) are then simply the $L$ momentum states
\begin{equation}
\Phi_m \approx \frac{1}{\sqrt{L}} \sum_{j=0}^{L-1} e^{i j m}\left |j \right\rangle.
\end{equation}
  
The matrix elements appearing in (\ref{linresponse}) are gently behaved as a function of system size. We easily find that  $\left\langle \Phi_{n}\right|b_{i+1}^{\dagger}b_{i+1}-b_{i}^{\dagger}b_{i}\left|\Phi_{m}\right\rangle = O(1)$ does not scale with $L$, while $\left\langle \Phi_{m}\right|H_{\rm SB}\left|\Phi_{n}\right\rangle \sim 1/\sqrt{L}$, since it is a sum of $L$ uncorrelated random variables. 
In contrast, the energy denominators $E_{m}-E_{n}$ are very small, since the degeneracy between the localized barrier configurations $|j\rangle$ is lifted only at the very high perturbative orders in $\lambda$. 
A naive estimate suggests that hybridization occurs via matrix elements of the order of $O(\lambda^{\rho_{b}L})$. However, considering that occasional local resonances (such as the triplet of intervals in Fig.~\ref{fig:1st_order_deg}
) help to connect distant states the actual matrix element between the various $|j\rangle$ will rather scale like $\lambda^{\alpha L}$ with $\alpha$ slightly smaller than $\rho_b$.
Accordingly, we expect that the eigenstate susceptibility to disorder grows exponentially with system size, as    
\begin{equation}
\frac{d\Delta \rho}{dW} \sim \left( \frac{c}{(\lambda/J)^{\alpha}}\right)^L, \quad \alpha \lesssim \rho_b,
\label{divergentsusc}
\end{equation}
with a constant $c = O(1)$ which depends on the details of the spectrum of fast particles, their density, the energy density $\epsilon$ etc.

The exponentially large response to infinitesimal disorder is
characteristic of genuine manybody systems. This contrasts with free
particles, which also localize in infinitesimal disorder, at least for
$d\leq 2$, but with a susceptibility $d\Delta\rho/dW$ that grows only as a
power law with system size.\footnote{While free particles in 3d are not be
sensitive to infinitesimal disorder, the generalization of our model to
3d, with  slow particles cutting links for fast particles, and a barrier
density such that $1-\rho_b$ is below the quantum percolation threshold,
is still expected to break translational invariance spontaneously.}

\begin{figure}
\includegraphics[width=\textwidth]{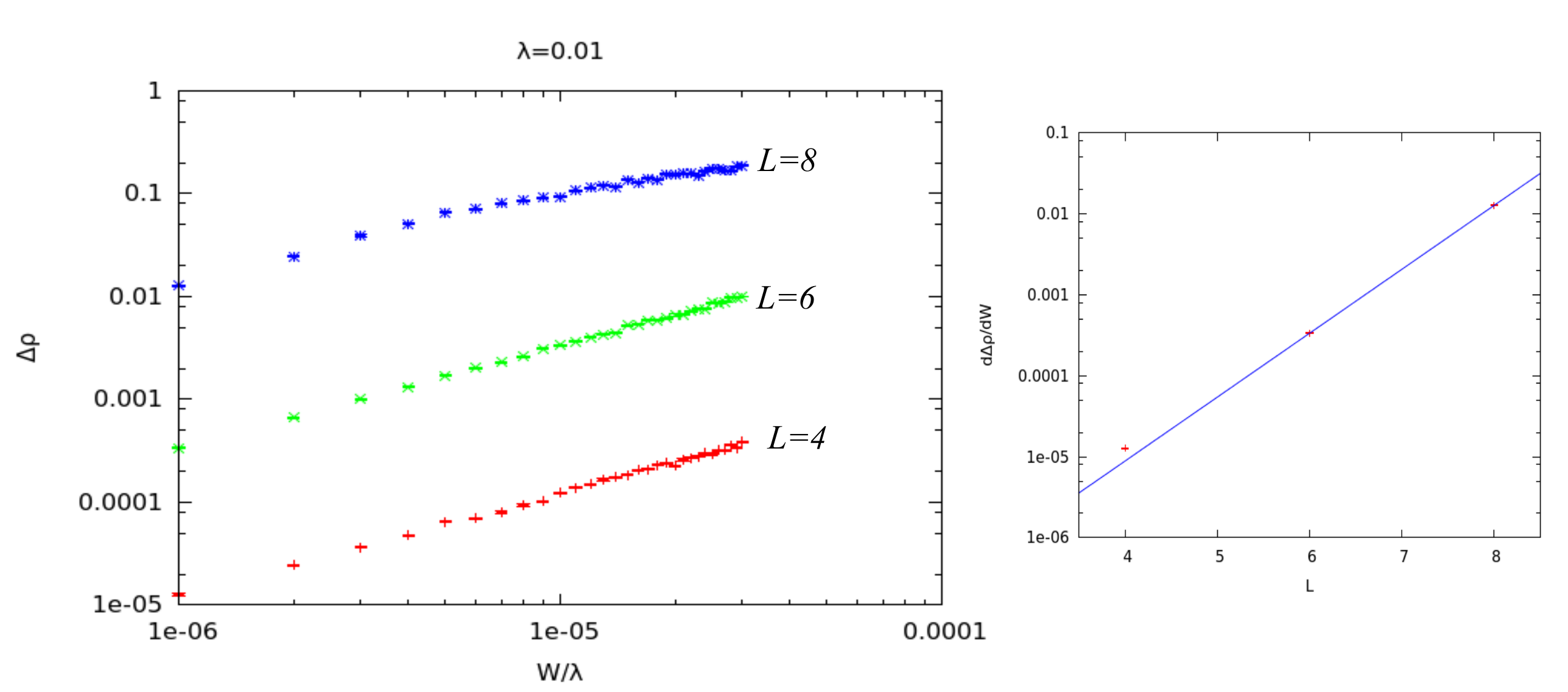}
\caption{Left: Disorder averaged spatial inhomogeneity $\Delta\rho(\epsilon=J/4)$ induced by a very weak disorder potential of strength $W\ll \lambda$ in a strongly localized system with barrier hopping $\lambda=0.01J$. The initial response is linear in $W$, with a susceptibility $d\Delta\rho/dW$ that diverges exponentially with system size. This demonstrates that in the thermodynamic limit the many-body eigenstates spontaneously break the translational symmetry, and thus violate the eigenstate thermalization hypothesis. This shows dynamically in the long time persistence of initial inhomogeneity and the absence of diffusion. -  Right: The fit of the disorder averaged susceptibility yields the exponential behavior $d\Delta \rho/dW \propto a^L$ with $a=6.2$ at $\lambda=0.01$.}
\label{fig:inhom}
\end{figure}

We have confirmed the above expectations numerically by computing the inhomogeneity $\Delta\rho$ in the presence of a very small disorder $W$.  We exactly diagonalized systems of sizes $L=4,6,8$ with a fixed
density of fast particles and barriers, $\rho_{a}=\rho_{b}=\frac{1}{2}$.
We averaged the results over 100 realizations of the disorder
and over a small energy window centered at energy density $\epsilon = E/L = J/4$. 
\footnote{In this way, we safely avoid the middle of the many-body energy band, $E=0$, which is highly degenerate at $\lambda=0$ because of the particle-hole symmetry in the sector of fast particles.} 
We have averaged $\overline{\Delta\rho}^\Phi$ over 10 eigenstates for $L=4,6$, and over 20 eigenstates for $L=8$. The results are shown in Fig.~\ref{fig:inhom} for $\lambda=\lambda_0=0.01J$.

It can be seen from the plot that the average susceptibility to disorder increases exponentially with the system size. Fitting the average susceptibility to the expected behavior (\ref{divergentsusc}) yields  the value $c/(\lambda_0/J)^\alpha \approx 6.2$.  From this we can obtain a rough estimate for the critical value of the barrier hopping, by assuming that Eq.~(\ref{divergentsusc}) holds approximately up to the delocalization transition. Since the susceptibility to disorder must stop growing exponentially with $L$ in the ergodic phase,  we may estimate the critical hopping from the requirement $c/(\lambda_c/J)^\alpha \approx 1$, which yields $\lambda_c \approx 
6.2^{1/\alpha} \lambda_0  \approx 0.4J$, where we approximated $\alpha\approx \rho_b$. 
This is consistent with the expectation that the delocalization or quantum glass transition takes place when the barrier hopping strength $\lambda$ becomes comparable to the hopping for the fast particles $J$. Nevertheless, it is interesting that our estimate suggests that quantum ergodicity requires a fairly large ratio of the hoppings (or in other words, of their effective masses). Thus the self-localization tendency of the quantum glass appears to be surprisingly robust. However, a more careful study of the delocalization transition and a more exhaustive study of the dependency on $\lambda$ is certainly needed to confirm this rough estimate.

\section{Discussion and conclusion}

\subsection{Potential application to magnets and quantum gases}
In the present study we chose to exemplify the mechanism that leads to quantum non-ergodicity in the simplest possible model, which may thus appear somewhat artificial. Our investigation is, however, inspired by realistic systems of disorder-free frustrated magnets~\cite{cepas}, whose low temperature dynamics were shown to be dominated by two types of elementary collective moves: a "fast" simultaneous flip of 6 spins along a ring, and a "slow" simultaneous flip of 10 spins along bigger rings. Under certain circumstances the fast flips are non-ergodic by themselves, and  may have fully localized dynamics, as long as the slow moves are frozen. Under such circumstances, and if the dynamics is essentially dominated by quantum tunneling rather than thermal activation, the essential ingredients for a quantum glass are indeed present. As long as the slow spin flips come with a small enough tunneling rate as compared to the fast ones, one may then expect such a quantum magnet to many-body localize in its self-generated disorder. It would be interesting to look for other realizations of similar physics, e.g., in 1d quantum spin chains. 

Another potential application of the presented ideas are strongly interacting quantum gases, see also Ref.~\cite{BoseHubbard}.
A recent numerical study of repulsive bosons, prepared in a strongly non-thermal, patterned initial condition, appeared to avoid thermalization and retain broken translational symmetry for surprisingly long times, exceeding the limits that could be simulated~\cite{BeccaFabrizio}. It is possible that the ideas presented in this work are relevant to understand the observed longevity of inhomogeneity in that system. Even though the  initial conditions chosen in Ref.~\cite{BeccaFabrizio} are periodic, it could be that their overlap with nearly periodic, delocalized eigenstates is nevertheless negligible as compared to the weight of localized eigenstates. 
    
\subsection{Enhanced delocalization at low $T$}
Our analysis is adapted to essentially random initial configurations, in which the barrier positions are uncorrelated. This is certainly a reasonable assumption for high energy densities in the initial state. However, if the system is prepared in an equilibrium configuration at very low temperature below the scale $J \rho_b^2$ (e.g., by weakly coupling the system to a bath for some time, and then switching the coupling off), one expects the fast particles to fall into the ground state levels confined between two neighboring barriers. To minimize the energy of the fast particles, the intervals between barriers must be rather homogeneous, and certainly far from being exponentially distributed, unlike what we assumed in Eq.~(\ref{expdist}). Under such circumstances, the probability of finding configurations that resonate at low orders of perturbation theory (similarly as in Fig.~\ref{fig:1st_order_deg}) increases significantly. One thus expects that already lower values of $\lambda$ will suffice to induce delocalization and restore ergodicity in the dynamics starting from such thermalized initial conditions. Viewing this from a different angle, one expects that at a fixed barrier hopping $\lambda$, a {\em decrease} of the temperature in the initial state renders the dynamics ergodic. With a grain of salt, this may be viewed as a cooling-induced melting of the quantum glass. However, one should bear in mind that the required temperature change and initial thermalization always requires the coupling to an external bath, which seems rather artificial.  

\subsection{On "integrability"}
In the model we considered, $H_0$ breaks the Hilbert space for the fast particles into small, closed intervals. Such a drastic form of non-ergodicity of $H_0$ seems not really necessary to arrive at our conclusion. In a 1d system even a mere modification of the fast hopping by the presence of a slow particle  should suffice to localize all fast particle states. In that case, $H_0$ will still be "integrable", that is, characterized by an extensive set of local conserved operators, the barrier positions, and the occupation numbers of states of fast particles in a given arrangement of slow particles. With appropriate modifications we expect that reasonings like given above still lead to the conclusion that for sufficiently small $\lambda$ the system is many-body localized, even though the threshold for delocalization might be significantly reduced.

Let us finally point out an interesting formal aspect of the  ideal quantum glass models we have analyzed. It should be emphasized once more that our analysis predicts many-body localization in the dynamics only for typical, i.e., sufficiently random initial conditions. This restriction is very mild, however, as it covers essentially all initial conditions up to a set that occupies an exponentially small fraction of phase space in the thermodynamic limit. Nevertheless, there exist long range correlated, nearly periodic initial conditions, for which our arguments may fail since system spanning hybridization occurs at low orders in perturbation theory, implying diffusion and transport. From the existence of these exceptional states, an interesting difference to many-body localized systems with quenched disorder seems to follow. For the latter,  it has been argued that an extensive set of mutually commuting, spatially localized, conserved operators should exist, whose eigenvalues fully characterize the many-body eigenstates.~\cite{conservedquantities1,conservedquantities2} However, in disorder-free systems, it is hard to see how such a set of local conservation laws could be compatible with the existence of rare delocalized states, as the former would seem to imply absence of thermalization in {\em any} initial state. 
Our preliminary attempts to construct such exponentially localized operators which commute with the Hamiltonian have indeed failed. 

This observation then raises the interesting question as to the precise relation between many-body localization and "integrability" in the sense of the existence of an extensive set of localized conserved operators. Does one imply the other, but not vice versa? Do systems with quenched disorder and those with self-generated disorder differ qualitatively in other respects? 
We hope to address these questions, as well as the nature of the many-body localization transition in systems with self-generated frustration in future work. 

After completion of this work, we became aware of Refs.~\cite{BoseHubbard, GrovelFisher}. In Ref.~\cite{BoseHubbard}  the idea of localization by self-generated (thermal) disorder is proposed, and it is shown rigorously that the thermal conductivity of a modified Bose-Hubbard model decays faster than any power of $\frac{1}{T}$ as $T\to \infty$. Related ideas of localization and lack of extensive entanglement in systems with heavy and light particles, but no quenched disorder, were recently discussed in Ref.~\cite{GrovelFisher}.

\subsubsection*{ACKNOWLEDGMENTS}
We thank B. L. Altshuler, G. Carleo, M. Fabrizio, L.B. Ioffe,  V. Kravtsov, A. Scardicchio and M. Wyart for useful discussions. We thank O. C\'epas for stimulating discussions on 2d compass models, which stimulated the present work. MM acknowledges KITP for hospitality during the initial stages of this work. This research was  supported in part by the National Science Foundation under Grant No. PHY11-25915.

\end{document}